\documentclass[aps,prd,superscriptaddress,nofootinbib,amsmath,amsfonts,preprintnumbers,groupedaddress,showpacs,8pt,english]{revtex4-1}
\usepackage{amsmath}
\usepackage{amssymb}
\usepackage{babel}
\usepackage{wrapfig}
\usepackage{cancel}

\usepackage{relsize,exscale}
\makeatletter
\usepackage{float}
\usepackage{array,multirow,graphicx}
\usepackage{dcolumn}
\usepackage{newlfont}
\usepackage{bm}
\usepackage[colorlinks,citecolor=blue,urlcolor=blue,linkcolor=blue]{hyperref}
\usepackage[figtopcap]{subfigure}
\usepackage{color}
\usepackage{float}
\usepackage{xcolor}

\newcommand{\am}[1]{{\bf\textcolor{magenta}{[Amare: #1]}}}

\newcommand{\rev}[1]{\textcolor{blue}{#1}}
\@ifundefined{color@BLUE}{\definecolor{BLUE}{rgb}{0,0,1}}{}

\usepackage{scalerel}
\usepackage{tikz}
\usetikzlibrary{svg.path}
\definecolor{orcidlogocol}{HTML}{A6CE39}
\tikzset{
  orcidlogo/.pic={
    \fill[orcidlogocol] svg{M256,128c0,70.7-57.3,128-128,128C57.3,256,0,198.7,0,128C0,57.3,57.3,0,128,0C198.7,0,256,57.3,256,128z};
    \fill[white] svg{M86.3,186.2H70.9V79.1h15.4v48.4V186.2z}
                 svg{M108.9,79.1h41.6c39.6,0,57,28.3,57,53.6c0,27.5-21.5,53.6-56.8,53.6h-41.8V79.1z M124.3,172.4h24.5c34.9,0,42.9-26.5,42.9-39.7c0-21.5-13.7-39.7-43.7-39.7h-23.7V172.4z}
                 svg{M88.7,56.8c0,5.5-4.5,10.1-10.1,10.1c-5.6,0-10.1-4.6-10.1-10.1c0-5.6,4.5-10.1,10.1-10.1C84.2,46.7,88.7,51.3,88.7,56.8z};}}
\newcommand\orcid[1]{\href{https://orcid.org/#1}{\mbox{\scalerel*{
\begin{tikzpicture}[yscale=-1,transform shape]
\pic{orcidlogo};
\end{tikzpicture}
}{|}}}}

\begin{document}
\title{Effective Constrained Scalar--Gauss--Bonnet Inflation Motivated by \(f(R,\mathcal{G})\) Gravity}
\author{G.G.L. Nashed$^{1,2}$}\email{nashed@bue.edu.eg}
\author{Sudan Hansraj$^{3}$}\email{hansrajs@ukzn.ac.za}
\author{Amare Abebe$^2$}\email{Amare.Abebe@nithecs.ac.za}
 
\affiliation{$^1$Centre for Theoretical Physics, The British University, Cairo, Egypt\\$^2$Centre for Space Research (CSR) North-West University, Potchefstroom 2520, ZA, South Africa\\ $^3$ Astrophysics Research Centre, Mathematics,  University of KwaZulu-Natal, Private Bag X54001, Durban 4000, South Africa.}
\date{\today}

\begin{abstract}
We develop an effective framework for inflation in a constrained scalar--Gauss--Bonnet theory motivated by a restricted sector of $f(R,\mathcal{G})$ gravity. Using unified parametrizations of the Hubble expansion rate and the Gauss--Bonnet coupling function within a generalized slow-roll formalism, we derive analytical expressions for the scalar spectral index $n_s$ and tensor-to-scalar ratio $r$, and study their dependence on the model parameters. We show that the Hubble parametrization primarily governs the scalar sector through the slow-roll parameter $\epsilon_1$, whereas the Gauss--Bonnet contribution, encoded in $\epsilon_4$, can significantly affect the scalar spectral tilt and suppress primordial tensor perturbations, naturally leading to very small values of the tensor-to-scalar ratio $r$. For the benchmark parameter set considered in this work, we obtain
$n_s \simeq 0.958$, and  $r \simeq 5.3\times10^{-4}$ with the tensor-to-scalar ratio remaining well below current observational upper bounds. These results demonstrate that the model can simultaneously produce a nearly scale-invariant scalar power spectrum and a highly suppressed tensor sector, compatible with current cosmological constraints. We further investigate the scalar perturbation structure of the exactly constrained theory, where the Lagrange multiplier constraint forces the lapse perturbation to vanish and, together with the gravitational momentum constraint, implies $\dot{\mathcal{R}}=0$, eliminating the propagating scalar degree of freedom at linear order. This exact result clarifies that the generalized slow-roll treatment should be interpreted as an effective softly constrained description. We also discuss perturbative stability conditions, including the positivity of the relevant kinetic coefficients and propagation speeds. Our analysis highlights the distinction between the exact constrained limit and its effective slow-roll realization, while showing that the effective constrained scalar--Gauss--Bonnet framework provides a flexible phenomenological description of inflation.
\end{abstract}
\maketitle

\section{Introduction}

The inflationary paradigm has become a cornerstone of modern cosmology, providing a compelling resolution to several fundamental shortcomings of the standard Big Bang framework, including the horizon, flatness, and monopole problems. In addition, inflation offers a natural mechanism for generating primordial quantum fluctuations that seed the formation of large-scale structures and leave observable imprints in the cosmic microwave background (CMB) anisotropies. These primordial perturbations are commonly characterized by two key observables: the scalar spectral index $n_s$ and the tensor-to-scalar ratio $r$, which serve as primary probes for distinguishing among competing inflationary scenarios \cite{Guth:1980zm,Linde:1981mu,Albrecht:1982wi,Mukhanov:1981xt,Mukhanov:1982nu,Liddle:2000cg,Bassett:2005xm,Baumann:2009ds,Lyth:1998xn,Linde:2007fr,Planck:2018vyg,Planck:2018jri,ACT:2020gnv,ACT:2023kun,Bitaj:2023kcg,Ellis:2025zrf,BICEP:2021xfz}.

Despite its remarkable success, the fundamental origin of inflation remains an open question. While canonical scalar--field models provide a simple and predictive framework, they often require fine-tuning of the potential and lack a direct connection to a more fundamental theory. This has motivated the exploration of alternative approaches based on modified theories of gravity, where inflation arises from geometric effects rather than an ad hoc scalar sector. Among these, $f(R)$ and $f(R,\mathcal{G})$ gravity theories have attracted considerable attention due to their ability to incorporate higher-curvature corrections naturally arising in quantum gravity and string-inspired frameworks \cite{Nojiri:2010wj,DeFelice:2010aj,Sotiriou:2008rp,Capozziello:2011et,Clifton:2011jh,Odintsov:2022qnn,Nojiri:2005vv,DeFelice:2009aj,DeLaurentis:2015fea,Odintsov:2020nwm,Nojiri:2020wmh,Capozziello:2021krv,Nojiri:2017ncd}.

In particular, the Gauss--Bonnet invariant $\mathcal{G}$ plays a central role in string-inspired effective actions, where it appears as a leading higher-curvature correction. When non-minimally coupled to scalar degrees of freedom, Gauss--Bonnet terms can significantly modify the inflationary dynamics and lead to distinct observational signatures. Recent studies have shown that the inclusion of Gauss--Bonnet couplings can reconcile otherwise excluded inflationary models with observational constraints, particularly by suppressing the tensor-to-scalar ratio without significantly affecting the scalar spectral index \cite{Kanti:1998jd,Satoh:2007gn,Guo:2010jr,Koh:2014bka,Odintsov:2018zhw,Odintsov:2020sqy,Nojiri:2017ncd,Oikonomou:2025ccs,Zhu:2025twm,Odintsov:2026cxz,Mudrunka:2025xcg}. In fact, recent analyses incorporating Atacama Cosmology Telescope (ACT) data indicate that Gauss--Bonnet couplings can alleviate tensions between theoretical predictions and observational measurements of $n_s$ and $r$ \cite{ACT:2023kun,Zhu:2025twm}.

However, generic higher-derivative gravity theories often suffer from Ostrogradsky instabilities, leading to ghost degrees of freedom that render the theory physically inconsistent.  The construction of perturbatively consistent formulations is therefore essential for ensuring theoretical viability.  One promising approach involves the introduction of auxiliary fields and Lagrange multiplier constraints, which effectively remove unwanted dynamical degrees of freedom while preserving the desirable features of higher-curvature corrections \cite{Woodard:2014iga,Langlois:2017dyl,Nojiri:2018ouv,Nojiri:2026hij,Horndeski:1974wa,Gleyzes:2014dya,Langlois:2017mdk,Langlois:2018dxi}. Such constrained higher-curvature constructions may be represented effectively within an Einstein--scalar--Gauss--Bonnet framework.

Within this framework, the inflationary observables depend on both the
background expansion history and the structure of the Gauss--Bonnet
coupling function. The scalar spectral index is mainly governed by the background Hubble evolution, while the Gauss--Bonnet sector provides subleading corrections through the generalized slow-roll hierarchy.  In contrast, the
tensor-to-scalar ratio is particularly sensitive to higher-curvature
corrections arising from the Gauss--Bonnet sector.

Within this setup, we derive the expressions for the scalar spectral index $n_s$ and the tensor-to-scalar ratio $r$ using a generalized slow-roll formalism and perform a numerical exploration of the parameter space. Our analysis reveals a nontrivial interplay between the background evolution and higher-curvature corrections. The Hubble parametrization
strongly influences the scalar sector through the slow-roll parameter $\epsilon_1$, while the Gauss--Bonnet-induced contribution
$\epsilon_4$ can provide important corrections to the scalar tilt. In contrast, the tensor sector is particularly sensitive to the
Gauss--Bonnet coupling, providing a natural mechanism for suppressing the tensor-to-scalar ratio. We confront our predictions with current observational constraints from Planck 2018, BICEP/Keck, and ACT data, and identify viable regions of parameter space consistent with $n_s \simeq 0.965$ and $r < 0.036$; the explicit benchmark constructed below, however, yields $n_s\simeq0.958$ (see Sec.~\ref{V}). In addition, we impose theoretical consistency conditions, including the absence of ghost and gradient instabilities in the tensor sector; a complete scalar no-ghost analysis (positivity of $Q_s$ and $c_s^2$) is deferred to future work, and within the slow-roll treatment the positivity of the auxiliary quantity $E(N)$ is used only as an approximate scalar-sector diagnostic. These results demonstrate that the proposed framework is both predictive and flexible, offering a consistent description of inflation within modified gravity.

The main novelty of the present work lies in the construction of an effective constrained scalar--Gauss--Bonnet inflationary framework equipped with a regularized Gauss--Bonnet coupling function. This structure allows a partial separation between the background dynamics governing the scalar sector and the higher-curvature corrections controlling the tensor sector. As a consequence, the model naturally realizes a strong suppression of primordial tensor modes while maintaining compatibility with current observational constraints on the scalar spectral index.

The structure of this paper is as follows. In Sec.~\ref{II}, we present the theoretical framework of the effective constrained $f(R,\mathcal{G})$ realization and derive the corresponding field equations. In Sec.~\ref{III}, we formulate the inflationary dynamics and compute the observable quantities using a generalized slow-roll approach. In Sec.~\ref{IV}, we specify the parametrizations of the Hubble parameter and Gauss--Bonnet coupling function. In Sec.~\ref{V}, we construct observationally viable solutions and analyze the parameter space. The numerical methodology and data analysis are described in Sec.~\ref{VI}, followed by a detailed discussion of the results in \rev{Sec.}~\ref{VII}. Finally, we summarize our findings and present concluding remarks in Sec.~\ref{VIII}.

\section{Theoretical Framework}\label{II}

We begin with an effective constrained scalar--Gauss--Bonnet representation of the \(f(R,\mathcal{G})\) gravitational theory. The action is written as
\begin{equation}
S=\int d^4x\sqrt{-g}\left[
\frac{1}{2\kappa^2}R
-\frac{1}{2}g^{\mu\nu}\partial_\mu\chi\partial_\nu\chi
-V(\chi)
+\lambda\left(g^{\mu\nu}\partial_\mu\chi\partial_\nu\chi+\mu^4\right)
+h(\chi)\mathcal{G}
+\mathcal{L}_m
\right],
\label{eq:action}
\end{equation}
where $g$ is the determinant of the metric tensor $g_{\mu\nu}$, $R$ is the
Ricci scalar, $\kappa^2=8\pi G_N$, and $G_N$ is Newton's gravitational constant.
The field $\chi$ denotes an auxiliary scalar degree of freedom, $V(\chi)$ is its
scalar potential, and $\lambda$ is a Lagrange multiplier. The constant $\mu$
has dimensions of mass and sets the normalization of the constraint imposed on
$\chi$. The function $h(\chi)$ denotes the Gauss--Bonnet coupling function, while
$\mathcal{L}_m$ is the matter Lagrangian density.

The Gauss--Bonnet invariant is defined as
\begin{equation}
\mathcal{G}
=
R^2
-4R_{\mu\nu}R^{\mu\nu}
+R_{\mu\nu\rho\sigma}R^{\mu\nu\rho\sigma},
\label{eq:GB_def}
\end{equation}
where $R_{\mu\nu}$ and $R_{\mu\nu\rho\sigma}$ are the Ricci and Riemann tensors,
respectively.
We emphasize that the action in Eq.~(\ref{eq:action}) should not be interpreted as
a completely arbitrary \(f(R,\mathcal{G})\) theory written in its
original higher-curvature form. Instead, it represents an effective
constrained scalar--Gauss--Bonnet realization motivated by a restricted
class of higher-curvature \(f(R,\mathcal{G})\) models. In particular, after the introduction of auxiliary scalar degrees of
freedom together with suitable Lagrange-multiplier constraints,
certain constrained higher-curvature constructions may be effectively
represented by an Einstein--scalar--Gauss--Bonnet description.

In the present work, the theory is interpreted as an effective constrained scalar--Gauss--Bonnet realization motivated by a restricted class of higher-curvature \(f(R,\mathcal{G})\) models. The Lagrange-multiplier constraint is introduced to suppress the additional higher-derivative degree of freedom associated with the constrained effective higher-curvature sector at the background level.   From a physical perspective, the constraint effectively restricts the dynamics of the auxiliary scalar sector, thereby reducing the impact of potentially pathological higher-derivative excitations inherited from the underlying higher-curvature theory.  The purpose of the present framework is therefore
to investigate the inflationary phenomenology and perturbative
stability properties of this constrained effective realization.

Variation of the action \eqref{eq:action} with respect to the Lagrange multiplier $\lambda$ yields the constraint
\begin{equation}
g^{\mu\nu}\partial_\mu\chi\partial_\nu\chi+\mu^4=0.
\label{eq:constraint}
\end{equation}
This Lagrange-multiplier constraint is introduced to suppress additional propagating degrees of freedom.

It is important to clarify the sense in which the term ``perturbatively consistent'' 
is used in the present work. The constraint in Eq.~(\ref{eq:constraint}) removes the unwanted higher-derivative degree of freedom at the level of the constrained background construction. Nevertheless, this condition by itself does not provide a complete demonstration of perturbative ghost freedom. A consistent stability analysis additionally requires the positivity of the kinetic coefficients and propagation speeds associated with the scalar and tensor perturbations. Accordingly, the term ``perturbatively consistent within the generalized slow-roll approximation'' is employed throughout this work only in an effective restricted sense, referring to the constrained realization supplemented by the perturbative stability conditions imposed below.

Variation with respect to the metric gives the modified Einstein equations
\begin{equation}
\frac{1}{\kappa^2}G_{\mu\nu}
=
T_{\mu\nu}^{(m)}
+
T_{\mu\nu}^{(\chi)}
+
T_{\mu\nu}^{(\mathcal{G})}, \qquad \mbox{where $G_{\mu\nu}$ is the Einstein tensor.}
\label{eq:Einstein}
\end{equation}
The scalar-field contribution is
\begin{equation}
T_{\mu\nu}^{(\chi)}
=
(1-2\lambda)\partial_\mu\chi\partial_\nu\chi
+
g_{\mu\nu}
\left[
\left(\lambda-\frac12\right)
\partial_\rho\chi\partial^\rho\chi
+\lambda\mu^4
-V(\chi)
\right],
\label{eq:Tchi}
\end{equation}
and the Gauss--Bonnet contribution is
\begin{align}
T_{\mu\nu}^{(\mathcal{G})}
=
-8
\Big[
R_{\mu\rho\nu\sigma}
+
R_{\rho\nu}g_{\sigma\mu}
-
R_{\rho\sigma}g_{\mu\nu}
-
R_{\mu\nu}g_{\sigma\rho}
+
R_{\mu\sigma}g_{\nu\rho}
+\frac12 R
\left(
g_{\mu\nu}g_{\sigma\rho}
-
g_{\mu\sigma}g_{\nu\rho}
\right)
\Big]
\nabla^\rho\nabla^\sigma h(\chi).
\label{eq:TGB}
\end{align}
It is worth noting that Eq.~(\ref{eq:TGB}) is the compact four-dimensional form of the metric variation of the scalar--Gauss--Bonnet term. Since the Gauss--Bonnet invariant is topological for a constant coupling, all algebraic terms proportional to \(h(\chi)\mathcal{G}\) cancel identically, and the contribution to the field equations depends only on second covariant derivatives of \(h(\chi)\).

Variation with respect to the scalar field $\chi$ yields
\begin{equation}
\nabla_\mu\left[(1-2\lambda)\nabla^\mu\chi\right]
+h_\chi(\chi)\mathcal{G}
-V_\chi(\chi)
=0,\qquad \mbox{where $h_\chi=d h/d\chi$ and $V_\chi=dV/d\chi$.}
\label{eq:scalar_eq}
\end{equation}
Equations \eqref{eq:Einstein}--\eqref{eq:scalar_eq}, together with the constraint
\eqref{eq:constraint}, form the complete set of field equations.

We consider a spatially flat Friedmann--Robertson--Walker spacetime,
\begin{equation}
ds^2=-dt^2+a(t)^2\left(dx^2+dy^2+dz^2\right), \quad \mbox{with Hubble parameter} \quad H=\frac{\dot a}{a}.
\label{eq:H}
\end{equation}
In this background, the Gauss--Bonnet invariant becomes
\begin{equation}
\mathcal{G}=24H^2\left(H^2+\dot H\right).
\label{eq:GB_FRW}
\end{equation}
The modified Friedmann equations read
\begin{equation}
\frac{3H^2}{\kappa^2}
=
\rho_m
+\left(\frac12-\lambda\right)\dot\chi^2
-\lambda\mu^4
+V(\chi)
-24H^3\dot h ,
\label{eq:F1}
\end{equation}
\begin{equation}
-\frac{1}{\kappa^2}\left(2\dot H+3H^2\right)
=
p_m
+\left(\frac12-\lambda\right)\dot\chi^2
+\lambda\mu^4
-V(\chi)
+8H^2\ddot h
+16H\left(\dot H+H^2\right)\dot h .
\label{eq:F2}
\end{equation}
The derivatives of the coupling function are
\begin{equation}
\dot h = \frac{dh}{dt}=h_\chi \dot\chi,
\qquad
\ddot h = \frac{d^2h}{dt^2}=h_{\chi\chi}\dot\chi^2 + h_\chi \ddot\chi.
\end{equation}
The constraint \eqref{eq:constraint} gives
\begin{equation}
-\dot\chi^2+\mu^4=0
\quad \Rightarrow \quad
\dot\chi=\pm \mu^2.
\end{equation}
Choosing the positive branch,
\begin{equation}
\chi=\mu^2 t+\chi_{0},\quad \mbox{and without loss of generality, we set $\chi_0=0$, so that} \quad \chi=\mu^2 t. \label{eq:chi_solution}
\end{equation}
Since we focus on inflation, we neglect ordinary matter:
\begin{equation}
\rho_m = p_m = 0.
\end{equation}
To construct explicit inflationary solutions, we later specify concrete parametrizations for the Hubble function and the Gauss--Bonnet coupling.

\section{Inflationary Observables}\label{III}

In the generalized Einstein--Gauss--Bonnet slow-roll formalism, the
inflationary dynamics receives contributions both from the background
expansion and from the higher-curvature Gauss--Bonnet sector.
Following the generalized slow-roll formulation of
Einstein--scalar--Gauss--Bonnet inflation
\cite{DeFelice:2010nf,DeFelice:2011uc}, and adapting the scalar kinetic
normalization to the constrained sector through the replacement
\(F\rightarrow 1-2\lambda\), we introduce\footnote{
In standard Einstein--Gauss--Bonnet inflationary theories, the
generalized slow-roll formalism introduces
\(
E = F + \frac{3Q_a^2}{2\dot\chi^2Q_T},
\)
where \(F\) denotes the effective scalar kinetic normalization,
\(Q_a=8H^2\dot h\), and \(Q_T=1+8H\dot h\)
\cite{DeFelice:2011uc}. Using
\(\dot h=Hh_N\), \(\dot\chi=H\chi_N\), and
\(F\rightarrow 1-2\lambda\), one obtains the effective expression
adopted in the present work.
}
the effective scalar kinetic quantity
\begin{equation}
E(N)=
(1-2\lambda)
+
\frac{96H^4h_N^2}
{\chi_N^2\left(1+8H^2h_N\right)} .
\label{Eeffective}
\end{equation}
The first term reflects the modified kinetic structure induced by the
Lagrange multiplier constraint, while the second term encodes the
higher-curvature corrections generated by the Gauss--Bonnet coupling.
The quantity \(E(N)\) is used here within the generalized slow-roll
hierarchy and should not be interpreted as the exact scalar kinetic
coefficient derived from the full quadratic scalar perturbation action.

The generalized slow-roll parameters are then written as\footnote{The effective generalized slow-roll hierarchy adopted in Eq.~(\ref{slowroll2}), together with the corresponding perturbative quantities in the constrained scalar--Gauss--Bonnet framework, is derived in Appendix~A.\\ The fourth slow-roll parameter is adopted from the generalized Einstein--Gauss--Bonnet perturbation formalism,
\[
\epsilon_4 \equiv \frac{\dot E}{2HE},
\]
where \(E\) denotes the effective scalar kinetic normalization entering the scalar perturbation sector. Using the e-folding variable \(dN=Hdt\), one has \(\dot E=HE_N\), which gives
\[
\epsilon_4=\frac{E_N}{2E}.
\]}
\begin{equation}
\epsilon_1 = -\frac{H_N}{H},
\quad
\epsilon_2=\epsilon_3=0,
\quad
\epsilon_4=\frac12\frac{E_N}{E},
\quad
\epsilon_5=
\frac{4H^2h_N}{1+8H^2h_N},
\quad
\epsilon_6=
\frac{4H\left(2H_Nh_N+Hh_{NN}\right)}
{1+8H^2h_N}.
\label{slowroll2}
\end{equation}
Here \(\epsilon_1\) is the usual Hubble slow-roll parameter. The
condition \(\epsilon_3=0\) follows from the constancy of the effective
gravitational coupling, while \(\epsilon_2=0\) follows from the
constraint-fixed scalar evolution. The parameters \(\epsilon_5\) and
\(\epsilon_6\) encode the Gauss--Bonnet corrections through derivatives
of the coupling function.

Throughout this work we employ reduced Planck units with
\(\kappa^2=1\), so that \(E(N)\) is dimensionless. Using the constraint
relation, one obtains
\begin{equation}
\chi_N
=
\frac{d\chi}{dN}
=
\frac{\dot\chi}{H}
=
\frac{\mu^2}{H},
\qquad
\dot\chi=\frac{d\chi}{dt}.
\label{chiN}
\end{equation}

The scalar spectral index is defined by
\begin{equation}
n_s-1
\equiv
\frac{d\ln\mathcal P_{\mathcal R}}
{d\ln k},
\label{nsdef}
\end{equation}
where \(\mathcal P_{\mathcal R}\) denotes the scalar curvature
perturbation power spectrum. In the present analysis, \(n_s\) is
evaluated within the generalized slow-roll hierarchy associated with
the background evolution and the Gauss--Bonnet sector.  
Here \(\mathcal P_{\mathcal R}(k)\) denotes the dimensionless scalar
curvature perturbation power spectrum, defined from the Fourier-space
two-point correlation function of the curvature perturbation
\(\mathcal R\) \cite{Liddle:2000cg}. The quantity
\(k\) represents the comoving wavenumber of the corresponding
perturbation mode, related to the comoving wavelength through
\(\lambda=2\pi/k\). In the inflationary regime, the scalar spectral
index characterizes the scale dependence of the curvature perturbation
spectrum around the horizon-crossing condition \(k=aH\)\footnote{Details of the evaluation of the scalar spectral index \(n_s\) within the effective generalized Einstein--Gauss--Bonnet slow-roll formalism are presented in Appendix~B.}.

The tensor-to-scalar ratio is evaluated using the effective
Einstein--Gauss--Bonnet slow-roll expression
\cite{DeFelice:2010nf,DeFelice:2011uc}
\begin{equation}
r
=
16
\left|
\frac{
\epsilon_1+\dfrac{Q_e}{4H}-\dfrac{Q_f}{4}
}{
(1+\frac{Q_b}{2})c_T^3
}
\right|,
\label{rdef}
\end{equation}
where
\begin{equation}
Q_b=16H^2h_N,
\qquad
Q_e=32H^2H_Nh_N,
\label{QbQe}
\end{equation}
and
\begin{equation}
Q_f
=
-16\left(
H^2h_{NN}
+
HH_Nh_N
-
H^2h_N
\right).
\label{Qf}
\end{equation}
The tensor propagation speed is correspondingly written as
\begin{equation}
c_T^2
=
1+\frac{Q_f}{1+\frac{Q_b}{2}}.
\label{cT}
\end{equation}

The expression for \(Q_f\) follows from rewriting the time-derivative
combination appearing in the Gauss--Bonnet tensor sector in terms of
\(N\)-derivatives. Since
\[
\dot h=Hh_N,
\]
one has
\[
\ddot h
=
\frac{d}{dt}(Hh_N)
=
\dot H h_N+H\dot h_N
=
HH_Nh_N+H^2h_{NN}.
\]
Therefore,
\[
\ddot h-H\dot h
=
H^2h_{NN}
+
HH_Nh_N
-
H^2h_N,
\]
which gives Eq.~(\ref{Qf}).

In the limit of a negligible Gauss--Bonnet coupling, the quantities
\(Q_b\), \(Q_e\), and \(Q_f\) vanish, \(c_T^2\to1\), and the standard
slow-roll expressions are recovered. We note, however, that the overall normalization of Eq.~(\ref{rdef}) should be checked against the conventions of Refs.~\cite{DeFelice:2010nf,DeFelice:2011uc}: a naive $Q_b,Q_e,Q_f\to0$, $c_T\to1$ limit of Eq.~(\ref{rdef}) gives $r\to4\epsilon_1$ rather than the canonical single-field value $r=16\epsilon_1$, indicating a prefactor that should be verified.

To ensure a regular tensor-sector evolution within the effective
description adopted here, we impose
\begin{equation}
1+\frac{Q_b}{2}>0,\qquad c_T^2>0.
\label{tensorstability}
\end{equation}
throughout the relevant inflationary regime and around the CMB pivot
scale. In addition, we require
\begin{equation}
E(N)>0,
\label{Econdition}
\end{equation}
as a consistency condition within the generalized slow-roll hierarchy.

We emphasize that the present analysis uses the effective
Einstein--Gauss--Bonnet slow-roll structure to investigate the
inflationary dynamics of the constrained scalar--Gauss--Bonnet
realization. A complete derivation of the full scalar perturbation
sector from the constrained action is carried out in Section~\ref{sec:scalar_perturbations}. We stress at the outset that the scalar observables computed below presuppose a \emph{softly} constrained realization, in which the constraint is understood to emerge dynamically from a steep confining potential so that the curvature perturbation remains weakly propagating. In the exact algebraic limit of the constraint, the analysis of Sec.~\ref{sec:scalar_perturbations} shows that the comoving curvature perturbation is frozen ($\dot{\mathcal R}=0$) and no scalar spectrum is generated; the values of $n_s$ reported here should accordingly be read as predictions of the effective soft-constraint description rather than of the exact constrained action in Eq.~(\ref{eq:action}).

Using the parametrizations
\begin{equation}
H(N)=H_0f(N;\theta),
\qquad
h(\chi)=g(\chi;\alpha),
\label{parametrization}
\end{equation}
all observables can be expressed in terms of the functions \(f\) and
\(g\) and their derivatives.

\section{Model Construction}\label{IV}

To investigate the phenomenological implications of the effective constrained scalar--Gauss--Bonnet realization of
$f(R,\mathcal{G})$ framework, it is necessary to specify the functional
forms of the Hubble parameter and the Gauss--Bonnet coupling function.
These choices determine the inflationary dynamics and directly influence
the observable quantities.

We parametrize the Hubble parameter as a function of the number of e-folds $N$:
\begin{equation}\label{27}
H(N) = H_0\, f(N),
\end{equation}
where $H_0$ is a constant and $f(N)$ is a dimensionless function describing
the background evolution.

In order to ensure a quasi-de Sitter expansion consistent with slow-roll
inflation, we restrict attention to forms of $f(N)$ that lead to a slowly
varying Hubble parameter, i.e., $|H_{N}/H| \ll 1$. Such parametrizations
naturally yield a nearly scale-invariant scalar spectrum.

The Gauss--Bonnet coupling function $h(\chi)$ governs the higher-curvature
corrections and plays a crucial role in shaping the tensor sector.
We consider several representative functional forms commonly studied
in the literature:

\begin{itemize}
\item Power-law coupling: \( h(\chi)=\alpha \chi^n \).
\item Exponential coupling: \( h(\chi)=\beta e^{\gamma \chi} \).
\item Hybrid coupling: \( h(\chi)=\delta \chi^m e^{\eta \chi} \).
\end{itemize}
In addition to these standard choices, we introduce a \textbf{regularized
Gauss--Bonnet coupling} of the form
\begin{equation}
h(\chi)=\gamma_h \frac{\chi^b}{1+\delta \chi^{\mathfrak{c}}},
\label{eq:regularized_h}
\end{equation}
which constitutes a central element of the present work. This form avoids
the rapid growth typically associated with power-law or exponential
couplings at large $\chi$, while preserving sufficient flexibility to
modify the tensor sector. As we show in the following sections, this
regularized structure allows for a significant suppression of the
tensor-to-scalar ratio while maintaining compatibility with the observed
values of the scalar spectral index.

Using the relation $\chi=\mu^2 t$, the coupling function can be expressed
in terms of cosmic time and, equivalently, in terms of the number of e-folds.
Consequently, all relevant quantities, including $h_{_\chi}(\chi)$ and $h_{_{\chi \chi}}(\chi)$,
can be written as functions of $N$.

The inflationary observables are computed by substituting the chosen
forms of $H(N)$ and $h(\chi)$ into the generalized slow-roll parameters
introduced in the previous section. The scalar spectral index $n_s$ and
tensor-to-scalar ratio $r$ are then evaluated at horizon crossing,
typically at $N \simeq 60$.

For the numerical analysis, the model parameters are selected such that
the slow-roll conditions are satisfied and the resulting predictions
fall within the observationally allowed ranges. Particular attention is
paid to the role of the regularized coupling \eqref{eq:regularized_h},
which proves essential in identifying viable regions of parameter space.

This approach enables a systematic exploration of how different
parametrizations of the Hubble expansion and Gauss--Bonnet coupling
affect the inflationary observables and their compatibility with
cosmological data.

\begin{figure}
\centering
\includegraphics[width=0.66\textwidth]{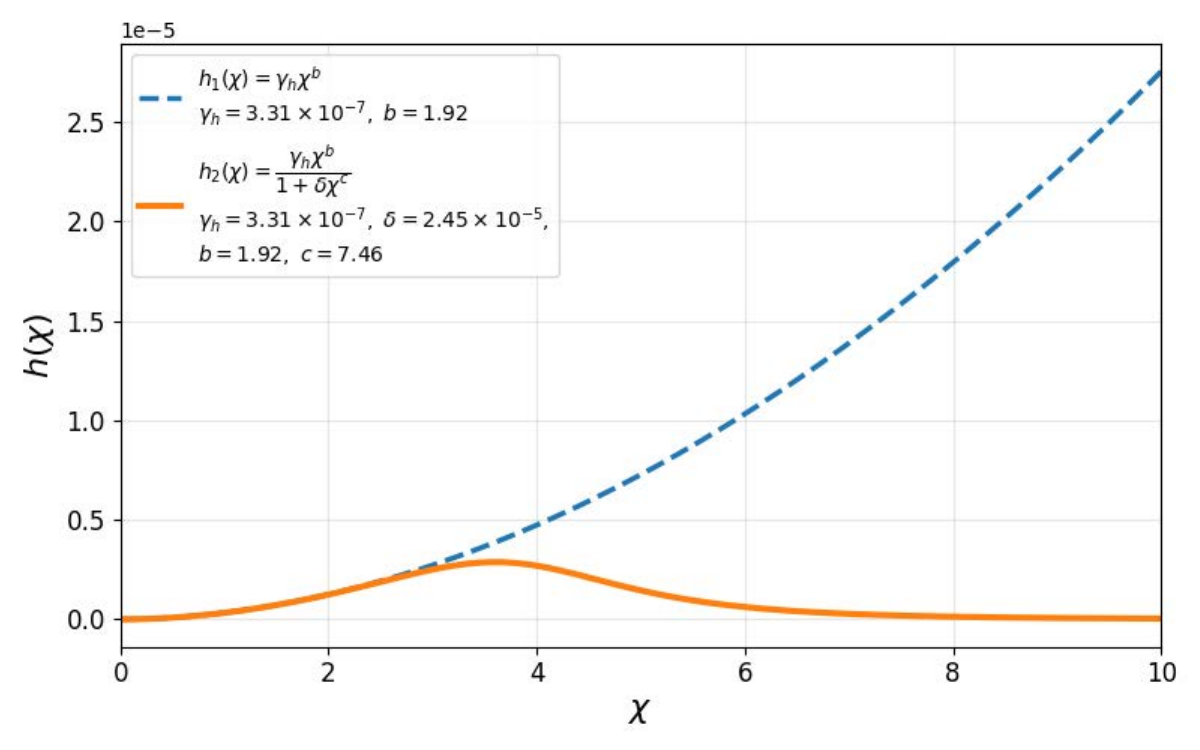}
\caption{Behavior of the regularized Gauss--Bonnet coupling function \(h(\chi)\). The coupling grows as a power law at small \(\chi\), while the denominator suppresses its growth at large \(\chi\), preventing uncontrolled higher-curvature corrections during inflation.}
\label{fig:hchi111}
\end{figure}

\section{Model Realization and Observationally Viable Parameter Space}\label{V}

In this section, we construct a phenomenologically viable realization of the effective constrained scalar--Gauss--Bonnet inflationary framework by specifying both the Hubble evolution and the Gauss--Bonnet coupling function. Our goal is to identify configurations that satisfy theoretical consistency conditions and are compatible with current observational constraints on the scalar spectral index $n_s$ and the tensor-to-scalar ratio $r$.

We consider an extended plateau-type parametrization of the Hubble parameter given by
\begin{equation}\label{38}
H(t) = H_0 \frac{1 + A e^{-pt}}{1 + B t^m},
\end{equation}
where $H_0$ sets the overall scale, while $A$, $p$, $B$, and $m$ control the early-time plateau behavior and the late-time exit from inflation.

At early times ($t \ll 1/p$), the Hubble parameter approaches a quasi-de Sitter phase,
\begin{equation}
H(t) \simeq H_0 (1 + A),
\end{equation}
while at late times the denominator induces a gradual departure from inflation, ensuring a natural exit.

We adopt the regularized coupling function
\begin{equation}\label{hchi}
h(\chi) = \gamma_h \frac{\chi^b}{1 + \delta \chi^\mathfrak{c}},
\end{equation}
which interpolates between a power-law regime at small $\chi$ and a suppressed coupling at large $\chi$. This behavior is essential to control the tensor sector and ensure compatibility with observational bounds.

A key aspect of the analysis is the consistent treatment of the number of e-folds,
\(
N(t)=\int_{t_i}^{t} H(t')\,dt'
\)
which provides the relation between cosmic time and observable quantities.
The total duration of inflation is then given by
\(
N_f=\int_{t_i}^{t_f} H(t)\,dt
\).
All observables are evaluated at horizon crossing, typically corresponding to $N \simeq 60$. In the numerical implementation, the evolution is computed in cosmic time, and the mapping to $N$ is obtained through numerical integration.

In practice, the Hubble parameter is expressed as a function of the e-folding number through the relation $H(N)=H(t(N))$, where $t(N)$ is obtained by numerically inverting $N(t)=\int H(t)\,dt$. All derivatives with respect to $N$ are computed after this transformation, ensuring full consistency with the slow-roll formalism.

The background dynamics can be reconstructed from the modified Friedmann
equations derived in Sec.~\ref{II}. Combining Eqs.~(\ref{eq:F1}) and (\ref{eq:F2}), and using the
constraint relation $\dot{\chi}^2=\mu^4$, one obtains a closed expression
for the Lagrange multiplier $\lambda$.

Adding the two Friedmann equations eliminates the potential $V(\chi)$ and yields
\begin{equation}
-\frac{2\dot H}{\kappa^2}
=
(1-2\lambda)\mu^4
+8H^2\ddot h
+8H(2\dot H-H^2)\dot h.
\end{equation}

Solving for $\lambda$, we obtain
\begin{equation}
\lambda
=
\frac{1}{2}
+
\frac{\dot H}{\kappa^2\mu^4}
+
\frac{4H^2\ddot h}{\mu^4}
+
\frac{4H(2\dot H-H^2)\dot h}{\mu^4}.
\label{eq:lambda_general}
\end{equation}

Using the relations
\begin{equation}
\dot h = h_\chi \dot\chi,
\quad
\ddot h = h_{\chi\chi}\dot\chi^2 + h_\chi \ddot\chi,\quad \mbox{together with the constraint} \quad
\dot\chi = \mu^2,
\qquad
\ddot\chi = 0,
\end{equation}
we obtain
\begin{equation}
\dot h = \mu^2 h_\chi,
\qquad
\ddot h = \mu^4 h_{\chi\chi}.
\end{equation}
Substituting into Eq.~\eqref{eq:lambda_general}, the Lagrange multiplier
takes the form
\begin{equation}
\lambda
=
\frac{1}{2}
+
\frac{\dot H}{\kappa^2\mu^4}
+
4H^2 h_{\chi\chi}
+
\frac{4H(2\dot H-H^2)}{\mu^2} h_\chi.
\label{eq:lambda_final}
\end{equation}

Using the relation $\chi = \mu^2 t$, the derivative with respect to the
number of e-folds is given by
\begin{equation}
\chi_{_N}(N) = \frac{d\chi}{dN} = \frac{\dot\chi}{H(N)} = \frac{\mu^2}{H(N)}.
\end{equation}

Accordingly, the derivative of the Gauss--Bonnet coupling function becomes
\begin{equation}
h_N(N) = \frac{dh(N)}{dN}=h_\chi \chi_{_N}(N) = h_\chi \frac{\mu^2}{H(N)}.
\end{equation}
The effective scalar kinetic normalization entering the perturbation
sector is represented here through the auxiliary quantity \(E(N)\),
which is employed within the generalized slow-roll hierarchy as an
effective indicator of scalar-sector stability.

The condition
\[
E(N) > 0
\]
is imposed as an approximate indicator of scalar-sector stability within the generalized slow-roll regime.

For consistency, the tensor-to-scalar ratio is computed as
\begin{equation}
r =16 \left|
\frac{\epsilon_1+\dfrac{Q_e}{4H}-\dfrac{Q_f}{4}}{(1+\frac{Q_b}{2})c_T^3}
\right|,\label{39}
\end{equation}
as defined in Sec.~\ref{III}.  In the limit of vanishing Gauss--Bonnet coupling, $h' = h'' = 0$, one has
$Q_b = Q_e = Q_f = 0$ and $c_T = 1$. Therefore, Eq.~(\ref{39}) reduces to
$r = 16\epsilon_1$, recovering the standard single-field slow-roll consistency relation.

We confront the model predictions with the latest observational constraints by adopting the benchmark parameter values
\begin{equation}
\begin{aligned}
&H_0 = 10.5377,\quad A = 0.0947,\quad p = 0.0153,\\
&B = 1.04\times10^{-5},\quad m = 1.62,\quad
\gamma_h = 2.36\times10^{-7},\\
&\delta = 1.82\times10^{-4},\quad
b = 4.893,\quad \mathfrak{c} = 4.077,\quad
\mu = 0.605 ,
\end{aligned}
\label{41}
\end{equation}
for which the inflationary observables are found to be
\begin{equation}
n_s \simeq 0.9581,\qquad r \simeq 5.3\times10^{-4}.
\label{42}
\end{equation}
These values are compatible with current observational bounds from the Planck and ACT collaborations, while predicting a very small tensor-to-scalar ratio.

\begin{figure}[H]
\centering
\subfigure[~Hubble parameter $H(N)$]{\label{fig:T1}\includegraphics[width=0.25\textwidth]{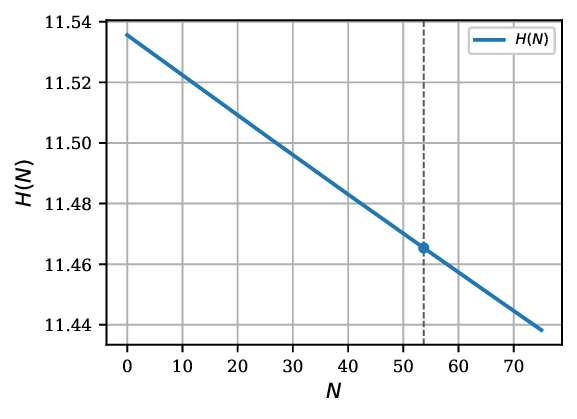}}
\subfigure[~effective kinetic function $E(N)$]{\label{fig:T2}\includegraphics[width=0.25\textwidth]{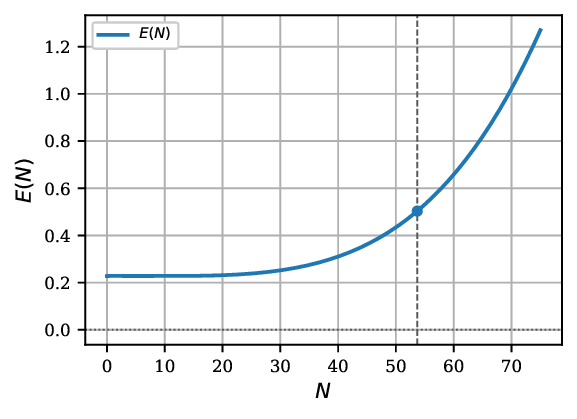}}
\subfigure[~$n_s$--$r$ plane with Planck constraints]{\label{fig:T3}\includegraphics[width=0.25\textwidth]{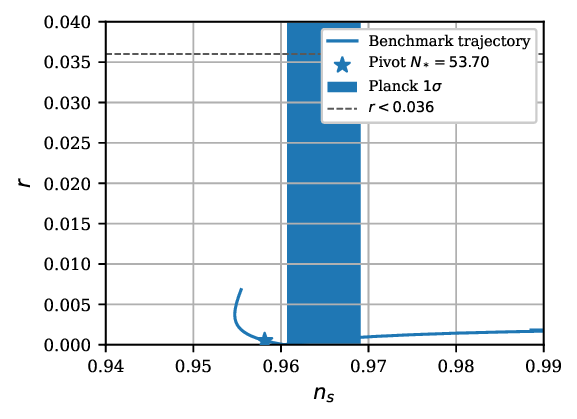}}
\subfigure[~tensor-to-scalar ratio $r(N)$]{\label{fig:T4}\includegraphics[width=0.25\textwidth]{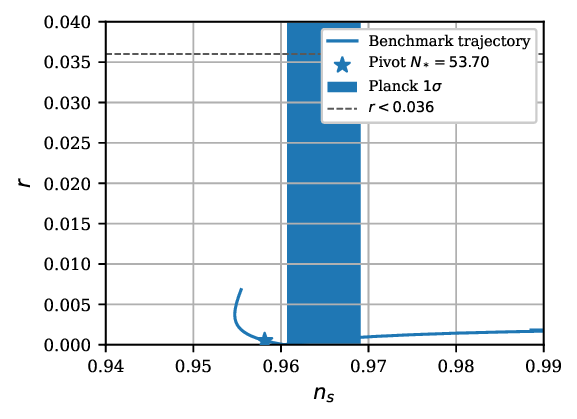}}
\subfigure[~tensor propagation speed $c_T^2(N)$]{\label{fig:T5}\includegraphics[width=0.25\textwidth]{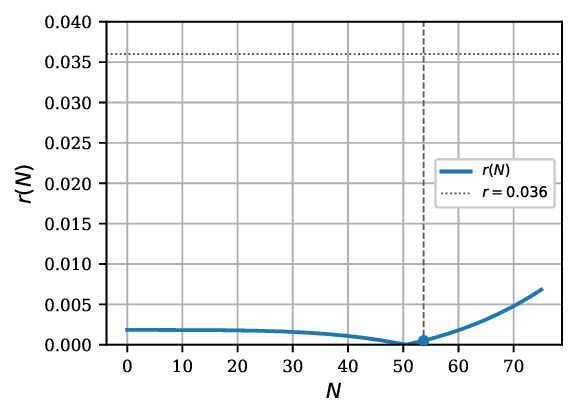}}
\subfigure[~scalar spectral index $n_s(N)$]{\label{fig:T6}\includegraphics[width=0.25\textwidth]{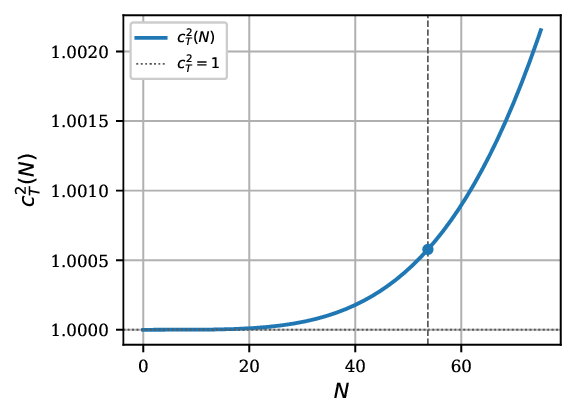}}
\subfigure[~parameter scan in the $n_s$--$r$ plane]{\label{fig:T7}\includegraphics[width=0.25\textwidth]{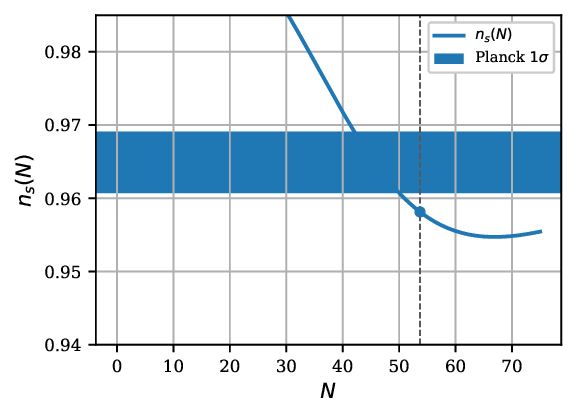}}
\caption{
Numerical evolution of the background and perturbation quantities, together with the parameter space analysis.
\subref{fig:T1} Hubble parameter $H(N)$, showing a quasi-de Sitter phase followed by a smooth exit from inflation.
\subref{fig:T2} Effective kinetic function E(N), remaining positive throughout the evolution and serving as an approximate indicator of scalar-sector stability within the generalized slow-roll regime.
\subref{fig:T3}  Distribution in the $n_s$--$r$ plane with Planck constraints, indicating observational viability.
\subref{fig:T4} Tensor-to-scalar ratio $r(N)$, demonstrating strong suppression during inflation.
\subref{fig:T5} Tensor propagation speed $c_T^2(N)$, remaining close to unity in agreement with gravitational wave constraints.
\subref{fig:T6} Scalar spectral index $n_s(N)$, evaluated during the inflationary evolution; the observational comparison is performed at the pivot scale
$N_\star\simeq53.70$. 
\subref{fig:T7}  Parameter scan in the $n_s$--$r$ plane, highlighting the restricted region of viable solutions.}
\label{fig:sectionV_results}
\end{figure}

The constraint scale $\mu$ enters the model through the relation
$\chi=\mu^{2}t$
and therefore affects the Gauss--Bonnet coupling function $h(\chi)$ and its derivatives. In all numerical calculations presented in this work, we adopt reduced Planck units $\kappa^{2}=1$ and set
$\mu=1$. With this choice, the benchmark parameter set given in Eq.~(\ref{benchmark}) reproduces all numerical results shown in Figs.~\ref{fig:sectionV_results} and \ref{Fig:4}.

It is worth noting that $\mu$ primarily determines the normalization of the auxiliary field $\chi$. Since the coupling function depends on the combination
$\chi=\mu^{2}t$, and $h(\chi)=\gamma_h\frac{\chi^{b}}{1+\delta\chi^{c}}$
a change in $\mu$ can be partially absorbed into a corresponding redefinition of the coupling parameters $\gamma_h$ and $\delta$. Consequently, $\mu$ mainly affects the overall normalization of the Gauss--Bonnet sector rather than introducing a qualitatively new inflationary behavior. For reproducibility, the benchmark value $\mu=1$ is now stated explicitly in the numerical setup.

In the numerical implementation we use reduced Planck units with $\kappa^2=1$, choose the positive branch $\dot{\chi}=\mu^2$, and set the initial time $t_i=0$. The e-fold variable is obtained from
\[
N(t)=\int_{t_i}^{t}H(t')\,dt',
\]
and the Hubble parameter is evaluated as \(H(N)=H(t(N))\) after numerically inverting this relation. The inflationary observables are evaluated at the pivot scale\footnote{Here $N_\star$ denotes the pivot e-folding number, namely the value of $N$ at which the observable quantities $n_s$ and $r$ are evaluated.} \(N_\star \simeq 53.70\), using the generalized slow-roll prescription described in Sec.~\ref{III}.

Additional parameter configurations within the viable region can also yield values closer to the central Planck estimate, namely
\[
n_s \sim 0.964 - 0.966,
\qquad
r \lesssim 10^{-3},
\]
while preserving the suppression of primordial tensor modes induced by the Gauss--Bonnet sector.
The predicted value of $r$ is comfortably consistent with current observational upper bounds from BICEP/Keck data, whereas the value of $n_s$ lies near the lower edge of the observationally allowed region. This indicates that the benchmark solution lies close to the observationally allowed region and motivates a more detailed exploration of the parameter space.

We caution that the present analysis constrains only the dimensionless observables $n_s$ and $r$, which are independent of the overall amplitude of the curvature spectrum. The scalar amplitude $\mathcal{A}_s\simeq2.1\times10^{-9}$ is not imposed here as a separate normalization condition, and the benchmark value $H_0=10.5377$ should be read as a convenient scale for the dimensionless evolution rather than as the physical inflationary Hubble rate. Matching $\mathcal{A}_s$ in standard single-field inflation fixes $H\sim10^{-5}M_{\mathrm{Pl}}$; imposing this normalization, together with the full scalar power-spectrum computation, is left to future work.

In addition, the model satisfies the consistency conditions
\begin{equation}
E(N) > 0, \qquad c_T^2(N) \simeq 1, \quad \mbox{supporting stability in the adopted slow-roll approximation.}
\end{equation}

\noindent
We clarify that the Lagrange-multiplier constraint in Eq.~(\ref{eq:constraint}) removes the unwanted higher-derivative mode at the level of the background construction, but by itself it does not constitute a complete proof of perturbative stability. A perturbatively consistent inflationary solution must also satisfy the positivity of the kinetic coefficients and sound speeds of the propagating scalar and tensor perturbations. Therefore, in addition to imposing the constraint, we require the conditions in Eqs.~\eqref{tensorstability} and (\ref{Econdition}) to be satisfied throughout the inflationary regime and, in particular, around the CMB pivot scale. Here $E(N)$ denotes the effective scalar kinetic function, while $Q_T$ and $c_T^2$ are the tensor kinetic coefficient and tensor propagation speed, respectively.  These conditions support perturbative consistency within the generalized slow-roll approximation, while a complete verification of scalar-sector stability requires the full evaluation of \(Q_s\) and \(c_s^2\).  Accordingly, the term ``perturbatively consistent within the generalized slow-roll approximation'' in the present work should be understood as referring to solutions that satisfy both the Lagrange-multiplier constraint and the perturbative stability conditions listed above.

The parameter space is explored through a numerical scan in which the model parameters are varied within physically motivated ranges. For each parameter set, the background evolution is computed, and the observables $n_s$ and $r$ are evaluated at \(N_\star \simeq 53.70\). The resulting distribution of points in the $n_s$--$r$ plane identifies the region compatible with observational constraints.
The end of inflation is identified through the condition
\[
\epsilon_1(N_f)=1,
\]
which marks the breakdown of the quasi-de Sitter phase and the onset of the post-inflationary evolution.

The resulting distribution in the  $n_s$--$r$ plane reveals a restricted region of solutions that are broadly compatible with current observational constraints, although some benchmark configurations remain close to the observational limits on the scalar spectral index.

The results demonstrate that achieving observational viability requires a simultaneous tuning of the Hubble parametrization and the Gauss--Bonnet coupling function. The extended plateau form of $H(t)$ plays a dominant role in determining the scalar spectral index, while the coupling function primarily affects the tensor sector.

The viable solutions occupy a narrow region of parameter space, highlighting the strong constraints imposed by current observational data and enhancing the predictive power of the model. Part of this narrowness reflects that the strong suppression of $r$ relies on a partial cancellation in the numerator of Eq.~(\ref{rdef}) between $\epsilon_1$ and the Gauss--Bonnet contributions $Q_e$ and $Q_f$. The predicted $r$ is therefore sensitive to the coupling parameters controlling this cancellation, and the very small value should be regarded as tuned rather than generic.

Figure~\ref{fig:sectionV_results} presents the numerical evolution of the background and perturbation quantities,
as well as the parameter space analysis. The Hubble parameter exhibits a quasi-de Sitter behavior followed by a smooth exit from inflation. The effective kinetic function remains positive throughout the evolution, supporting the perturbative consistency of the scalar sector within the generalized slow-roll approximation adopted here. The scalar spectral index remains close to the observationally favored region near the pivot scale, while the tensor-to-scalar ratio is significantly suppressed. Finally, the distribution in the $n_s$--$r$ plane shows that the model predictions occupy a region broadly compatible with current observational constraints, supporting the phenomenological viability of the framework.

\section{Methodology}\label{VI}

The phenomenological viability of the model is assessed through a numerical exploration of the parameter space. The analysis is performed by sampling the model parameters within physically motivated ranges and computing the corresponding inflationary observables.

For each parameter set, the background evolution is determined by specifying the Hubble parameter $H(N)$ and the Gauss--Bonnet coupling function $h(\chi)$. Using the constraint relation $\chi = \mu^2 t$ and the definition of the number of e-folds, all quantities are expressed in terms of $N$. In particular, one obtains
\begin{equation}
\chi_{_N}(N) = \frac{d\chi}{dN} = \frac{\mu^2}{H(N)}.
\end{equation}
The Gauss--Bonnet coupling function is then expressed as \(h(N)=h(\chi(N))\), and all derivatives, such as \(H_N(N)\), \(h_N(N)\), and \(h_{NN}(N)\), are computed with respect to \(N\).

The generalized slow-roll parameters are evaluated using expressions from the previous section, and the scalar spectral index $n_s$ and tensor-to-scalar ratio $r$ are computed at horizon crossing, typically at $N \simeq 50\text{--}60$. The numerical evaluation confirms that the scalar spectral index remains close to the observationally preferred range near the pivot scale. A complete numerical evaluation of the full scalar kinetic coefficient \(Q_s\) and scalar sound speed \(c_s^2\) is deferred to future work. In the present analysis, the positivity of the auxiliary effective kinetic quantity \(E(N)\) is used as an approximate indicator of scalar-sector stability within the generalized slow-roll regime.

The numerical implementation is carried out by performing a grid-based scan over the parameter space. For each parameter set, the background evolution and inflationary observables are computed, allowing us to construct the distribution of predictions in the $n_s$--$r$ plane and identify regions compatible with observational constraints.

The theoretical predictions are compared with current cosmic microwave background (CMB) observations. In particular, we consider constraints on the scalar spectral index and tensor-to-scalar ratio from:

\begin{itemize}
\item \textbf{Planck 2018 data release} \cite{Planck:2018vyg}, providing precise measurements of temperature and polarization anisotropies.
\item \textbf{Atacama Cosmology Telescope (ACT) DR6 data} \cite{ACT:2023kun}, offering complementary high-resolution observations.
\end{itemize}

These datasets impose the constraints
\begin{equation}
n_s = 0.9649 \pm 0.0042, \qquad r < 0.036, \qquad \mbox{which are used to assess the viability of the model.}
\end{equation}

To explore the dependence of the inflationary observables on the model parameters, we perform a systematic scan over the parameter space. Each parameter is varied within a predefined physically motivated range, and for each configuration, the observables are computed at horizon crossing.
For each parameter configuration, the solution was retained only if the following conditions were satisfied around the pivot scale:
\[
\epsilon_1 \ll 1, \qquad
E(N) > 0, \qquad
c_T^2(N) > 0,
\]
together with the observational requirements
\[
0.95 \lesssim n_s \lesssim 0.975,
\qquad
r < 0.036.
\]
Configurations violating any of these conditions were excluded from the viable parameter space.

For the numerical scan, the parameters were varied within the following representative ranges:
\[
\begin{aligned}
&H_0 \in [5,20], \qquad
A \in [0,0.5], \qquad
p \in [10^{-3},10^{-1}],\qquad B \in [10^{-7},10^{-3}],\qquad m \in [0,2,5],\\
&
 \gamma_h \in [10^{-9},10^{-5}], \qquad
\delta \in [10^{-7},10^{-3}],\qquad 
b \in [1,5], \qquad
{\mathfrak{c}} \in [1,10].
\end{aligned}
\]
These intervals were selected to ensure quasi-de Sitter evolution, perturbative consistency, and compatibility with observational constraints.

Although the present approach does not involve a full Bayesian
statistical analysis or Markov Chain Monte Carlo parameter estimation,
it nevertheless provides a robust qualitative exploration of the viable
parameter space and its compatibility with current observational
constraints.

\section{Results}\label{VII}

The quantities displayed in Figs.~\ref{fig:hchi111} and \ref{fig:sectionV_results} represent the complete numerical evolution of the inflationary observables during the slow-roll era, rather than only their values evaluated at the CMB pivot scale. Consequently, the scalar spectral index $n_s(N)$ may temporarily depart from the observationally preferred interval during certain stages of the evolution. The physically relevant predictions, however, are those computed at horizon crossing, typically around $N \simeq 50$--$60$. The benchmark values quoted throughout the text correspond to this pivot-scale evaluation. Consistency between the numerical results and the plotted quantities has been carefully verified.

In this section, we discuss the main outcomes of our analysis, with particular emphasis on how the inflationary observables depend on the chosen Hubble parametrization and on the form of the Gauss--Bonnet coupling function. The observables are evaluated numerically at horizon crossing and compared with current CMB constraints \cite{Planck:2018jri,BICEP2:2018kqh,ACT:2020gnv}.

The choice of the Hubble parametrization $H(t)$ has a significant impact on the inflationary predictions, especially on the scalar spectral index $n_s$. This reflects the well-known sensitivity of scalar perturbations to the background expansion history. In the present Einstein--scalar--Gauss--Bonnet framework, additional corrections also arise from the Gauss--Bonnet contribution encoded in the slow-roll parameter $\epsilon_4$.

The Hubble parameter enters directly into the first slow-roll parameter through Eq.~\eqref{slowroll2} which measures the departure from an exact de Sitter phase. Different functional forms of \(H(t)\) therefore modify \(\epsilon_1\) and consequently affect the scalar spectral index through the curvature perturbation spectrum discussed in Sec.~\ref{III}.

Our numerical analysis indicates that plateau-like parametrizations of the Hubble rate, such as the form introduced in Eq.~(\ref{27}), naturally produce values of $n_s$ consistent with observational bounds, namely $n_s \simeq 0.96$ \cite{Planck:2018jri}. On the other hand, rapidly varying or steeper Hubble functions generally lead to spectra that are either excessively red-tilted or blue-tilted and are therefore disfavored observationally.

These results underline the central role of the background expansion in determining the scalar sector, while the Gauss--Bonnet-induced contribution $\epsilon_4$ provides additional corrections to the scalar tilt \cite{Liddle:2000cg,Martin:2013tda}. This behavior can be understood from the structure of the generalized slow-roll hierarchy. The scalar spectral index depends primarily on the first slow-roll parameter \(\epsilon_1\), which directly measures the deviation from quasi-de Sitter expansion. Plateau-like Hubble evolutions keep \(\epsilon_1\) small and slowly varying over a large number of e-folds, thereby naturally generating an approximately scale-invariant scalar spectrum. In contrast, the Gauss--Bonnet sector contributes mainly through subleading corrections associated with \(\epsilon_4\), so its impact on the scalar tilt is comparatively moderate.

The Gauss--Bonnet coupling function $h(\chi)$ mainly influences the tensor sector, particularly the tensor-to-scalar ratio $r$. This dependence appears through the quantities $Q_b$, $Q_e$, and $Q_f$, which contain derivatives of $h(\chi)$, as shown in Eqs.~(\ref{QbQe})--(\ref{cT}).

Unlike minimally coupled inflationary scenarios, where $r$ is primarily controlled by the first slow-roll parameter, the inclusion of the Gauss--Bonnet term introduces additional contributions capable of either suppressing or enhancing tensor perturbations.

We find that the regularized coupling function $h(\chi)$ introduced in Eq.~\eqref{hchi} is especially effective in suppressing the tensor amplitude, yielding values as low as $r \sim 10^{-4}$ while preserving compatibility with the observed scalar spectral index. Physically, the suppression of the tensor sector originates from the derivative structure of the Gauss--Bonnet coupling, which modifies the effective tensor kinetic terms through the quantities \(Q_b\), \(Q_e\), and \(Q_f\). The regularized form of the coupling prevents an uncontrolled growth of these corrections at large field values while still allowing sufficiently large derivative contributions during the inflationary era. As a consequence, the tensor amplitude can be substantially reduced without inducing large distortions in the scalar sector. This mechanism explains why the model naturally predicts very small values of the tensor-to-scalar ratio while maintaining a scalar spectral index close to the observationally preferred region.

Such a suppression mechanism is in agreement with previous studies of Gauss--Bonnet inflation, where non-minimal couplings were shown to decrease the tensor amplitude \cite{Guo:2009uk,Koh:2014bka}. Furthermore, the strong dependence of $r$ on the structure of $h(\chi)$ offers a useful way to distinguish between different coupling choices.

By scanning the parameter space, we identify several configurations that simultaneously satisfy observational constraints on $n_s$ and $r$ while remaining theoretically consistent at the level of the tensor sector and the $E(N)>0$ diagnostic, with a complete scalar no-ghost analysis deferred to future work.

A representative benchmark model, defined in Eq.~(\ref{38}), gives $n_s$ and $r$  compatible with present observational bounds, although the scalar spectral index lies close to the lower edge of the Planck-allowed region \cite{Planck:2018jri,BICEP:2021xfz}.

The evolution of the background and perturbation quantities is shown in Fig.~\ref{fig:sectionV_results}. In particular,
\begin{itemize}
\item the Hubble parameter exhibits a quasi-de Sitter plateau followed by a smooth exit from inflation,
\item the effective kinetic function \(E(N)\) remains positive during the entire evolution, supporting perturbative consistency within the generalized slow-roll approximation adopted here,
\item the tensor propagation speed satisfies $c_T^2 \simeq 1$, consistent with gravitational-wave constraints \cite{LIGOScientific:2017vwq}.
\end{itemize}

In addition, the parameter scan presented in Fig.~\ref{fig:sectionV_results}\subref{fig:T7} shows that viable solutions occupy only a relatively restricted region in the $n_s$--$r$ plane. This suggests that the model is predictive and strongly constrained by current observations.

Overall, the analysis shows that observational viability emerges from a nontrivial interplay between the Hubble parametrization and the Gauss--Bonnet coupling function. The background evolution primarily controls the scalar spectral index through $\epsilon_1$, while the Gauss--Bonnet contribution $\epsilon_4$ introduces additional corrections to the scalar sector. In contrast, the tensor-to-scalar ratio is highly sensitive to the structure of the coupling function.

\begin{figure}[H]
\centering
\includegraphics[scale=0.66]{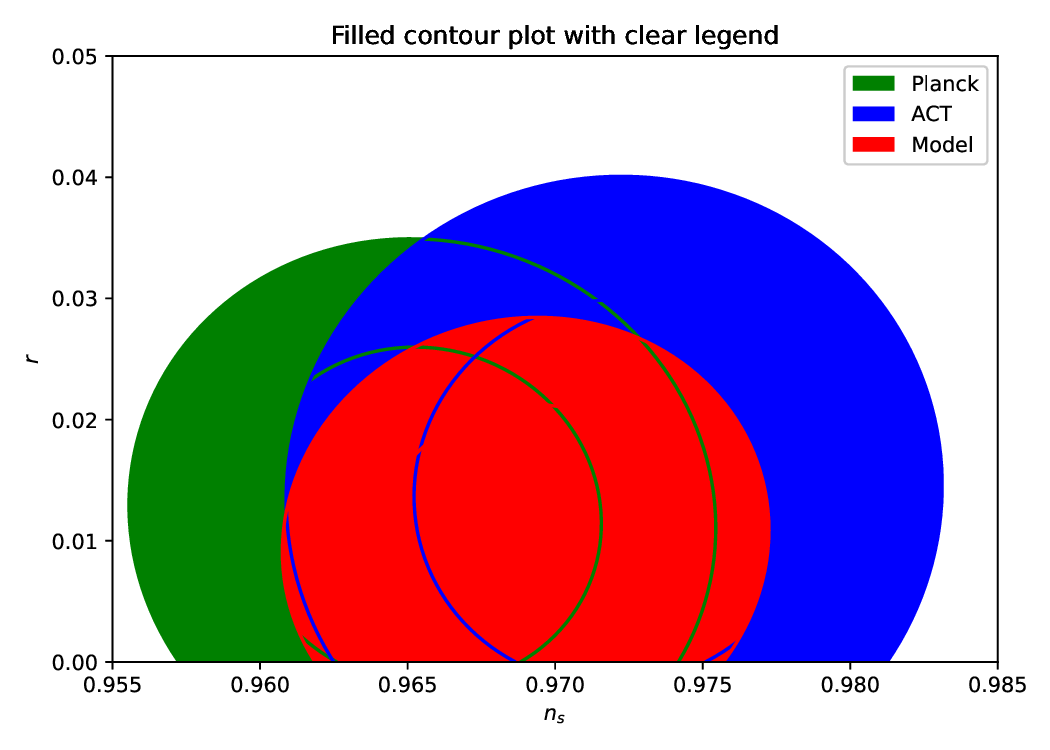}
\caption{
Filled contour plot in the $n_s$--$r$ plane displaying observational constraints together with the predictions of the present model.
The green and blue regions correspond to the confidence contours from Planck and ACT data, respectively, while the red region represents the allowed parameter space of the model.
The overlap between these regions indicates compatibility with current observational constraints.
}
\label{Fig:4}
\end{figure}

The observational consistency of the model is further illustrated in Fig.~\ref{Fig:4}, where the predicted parameter space is compared with the Planck and ACT confidence regions in the $n_s$--$r$ plane. The overlap confirms that the model remains compatible with present cosmological observations.

For completeness, Table~\ref{tab:comparison} summarizes the predictions of the present scenario and compares them with several representative inflationary models.

\begin{table}[H]
\centering
\begin{tabular}{lccc}
\hline
\textbf{Model} & \textbf{$n_s$} & \textbf{$r$} & \textbf{Remarks} \\
\hline
Standard single-field inflation
& $\sim 0.965$
& $\sim 10^{-2}$
& Typical slow-roll models \cite{Planck:2018jri,Tsujikawa:2013ila} \\

Gauss--Bonnet inflation (generic)
& $\sim 0.96$
& $\sim 10^{-3}$
& Tensor suppression from coupling \cite{Guo:2009uk,Koh:2014bka,Planck:2018jri} \\

Plateau inflation models
& $\sim 0.965$
& $\lesssim 10^{-3}$
& Good agreement with Planck \cite{Planck:2018jri,Starobinsky:1980te} \\

\textbf{Present model}
& $\mathbf{n_s \simeq 0.958}$  
& $\mathbf{r \simeq 5.3\times10^{-4}}$
& Suppressed tensor sector with \\
&&& observationally viable parameter space \\
\hline
\end{tabular}

\caption{
Comparison of inflationary observables for several representative inflationary scenarios.
The quoted ranges are indicative values reported in the literature and compatible with current CMB constraints.
}
\label{tab:comparison}
\end{table}

\section{Scalar perturbation structure of the exactly constrained theory and its relation to the effective slow-roll description}
\label{sec:scalar_perturbations}

In the preceding sections, the inflationary observables and the generalized slow-roll parameters were derived under the phenomenological assumption of a dynamical scalar degree of freedom. Such a treatment effectively assumes a ``soft" constraint, where the field dynamics are heavily restricted but the kinetic term retains enough freedom to allow the propagation of scalar perturbations, a scenario typically realized when the constraint emerges dynamically from a steep confining potential. However, it is interesting to examine the perturbation structure in the theoretical limit where the constraint $g^{\mu\nu}\partial_\mu\chi\partial_\nu\chi + \mu^4 = 0$ is enforced as an exact algebraic identity via a Lagrange multiplier. In this rigid limit, the nature of the physical degrees of freedom changes significantly. Consequently we carry out the scalar perturbation analysis that was deferred in Sec.~III. We show that the exact Lagrange‑multiplier constraint, combined with the gravitational momentum constraint, removes the propagating scalar degree of freedom from the theory at linear order. This result is exact, holds independently of the slow‑roll approximation, and clarifies \am{that} the constrained framework should be interpreted as defining a restricted sector of \(f(R,\mathcal{G})\) gravity where scalar perturbations do not contribute at linear order. 

\subsection*{Lapse perturbation in the comoving gauge}

We work in the comoving gauge \(\delta \chi = 0\) with the ADM decomposition
\[
\mathrm{d}s^{2} = -(1 + \alpha)^{2}\mathrm{d}t^{2} + 2a\partial_{t}\psi \,\mathrm{d}t\mathrm{d}x^{i} + a^{2}e^{2\mathcal{R}}\delta_{ij}\mathrm{d}x^{i}\mathrm{d}x^{j},
\]
where \(\alpha\) is the lapse perturbation, \(\psi\) the scalar shift, and \(\mathcal{R}\) the comoving curvature perturbation.

The Lagrange‑multiplier constraint given by Eq.~\eqref{eq:constraint}, \(g^{\mu \nu}\partial_{\mu}\chi \partial_{\nu}\chi +\mu^{4} = 0\), is evaluated in the comoving gauge. Since \(\delta \chi = 0\) the only contribution is
\[
g^{00}\dot{\chi}^{2} + \mu^{4} = 0.
\]
At zeroth order the background relation \(-\dot{\chi}^{2} + \mu^{4} = 0\) is satisfied identically. Expanding \(g^{00} = -(1 + \alpha)^{-2} = -(1 - 2\alpha) + \mathcal{O}(\alpha^{2})\) and linearizing gives
\[
2\alpha \dot{\chi}^{2} = 0 \quad \Longrightarrow \quad \alpha = 0, \quad \mbox{since \(\dot{\chi} = \mu^{2}\neq 0\).}
\]
Two features of this result deserve emphasis. First, the shift \(\psi\) does not enter because \(\partial_{i}\chi = 0\) in the comoving gauge; consequently \(\alpha = 0\) is determined by the constraint alone, without reference to the gravitational field equations. Second, the Lagrange‑multiplier perturbation \(\delta \lambda\) does not appear at this order: the first‑order variation of the constraint term \(\lambda (g^{\mu \nu}\partial_{\mu}\chi \partial_{\nu}\chi +\mu^{4})\) yields \(\delta \lambda \times 0 + \bar{\lambda}\times 2\alpha \mu^{4} = 0\), which is satisfied for any \(\delta \lambda\) once \(\alpha = 0\) is imposed. Thus \(\delta \lambda\) remains undetermined by the constraint at first order and plays no role in the scalar perturbation dynamics derived below.

\subsection*{Momentum constraint and the fate of the scalar mode}

The second‑order action for scalar perturbations in the general Horndeski framework  to which the Einstein-scalar-Gauss-Bonnet sector belongs after the standard field redefinition \cite{DeFelice:2010nf} - takes the form \cite{DeFelice:2011uc}
\[
S_{s}^{(2)} = \int \mathrm{d}t\mathrm{d}^{3}x\,a^{3}\left[3w_{1}\dot{\mathcal{R}}^{2} + \frac{1}{a^{2}}\bigl(2w_{1}\dot{\mathcal{R}} -w_{2}\alpha \bigr)\partial^{2}\psi +3w_{2}\alpha \dot{\mathcal{R}} +\frac{1}{3} w_{3}\alpha^{2} - \frac{2w_{1}}{a^{2}}\alpha \partial^{2}\mathcal{R} + \frac{w_{4}}{a^{2}} (\partial \mathcal{R})^{2}\right],
\]
where \(w_{1}-w_{4}\) are background‑dependent coefficients defined in Eqs.~(18)–(21) of Ref.~\cite{DeFelice:2011uc}. For the Einstein‑Gauss‑Bonnet sector with \(\kappa^{2}=1\), the Horndeski mapping yields \cite{DeFelice:2011uc,Kobayashi:2019hrl}
\[
w_{1} = 1 + 8H\dot{h},
\]
which is strictly positive during inflation and coincides with the tensor kinetic coefficient, \(Q_{T} = w_{1} = 1+8H\dot h\), consistently with the definition of \(Q_T\) given in Appendix~\ref{appA}.

The momentum constraint is obtained by varying the action with respect to the shift \(\psi\). Since \(\psi\) enters only through \((2w_{1}\dot{\mathcal{R}} - w_{2}\alpha)\partial^{2}\psi\), the variation yields \cite{DeFelice:2011uc}
\begin{equation}
2w_{1}\dot{\mathcal{R}} -w_{2}\alpha = 0. \label{eq:momentum}
\end{equation}
In the unconstrained theory, Eq.~\eqref{eq:momentum} determines \(\alpha = (2w_{1}/w_{2})\dot{\mathcal{R}}\), and the curvature perturbation \(\mathcal{R}\) remains dynamical. In the exactly constrained theory, however, \(\alpha = 0\) is imposed independently by the Lagrange‑multiplier condition. Substituting into the momentum constraint gives
\[
2w_{1}\dot{\mathcal{R}} = 0\quad \Longrightarrow \quad \dot{\mathcal{R}} = 0,
\]
since \(w_{1}\neq 0\). The comoving curvature perturbation is frozen: there is no propagating scalar degree of freedom in the exactly constrained theory. We stress that the Lagrange‑multiplier perturbation \(\delta \lambda\) cannot modify this conclusion. In the comoving gauge with \(\delta \chi = 0\), the constraint \(g^{\mu \nu}\partial_{\mu}\chi \partial_{\nu}\chi +\mu^{4}=0\) depends only on \(g^{00}\) and not on the shift \(\psi\); consequently \(\delta \lambda\) does not appear in the momentum constraint, and the result \(\dot{\mathcal{R}}=0\) is unaffected.

The result \(\dot{\mathcal{R}}=0\) has several important implications.
As demonstrated, enforcing the constraint rigidly dictates that the lapse perturbation vanishes in the comoving gauge ($\alpha = 0$), which through the momentum constraint strictly requires $\dot{\mathcal{R}} = 0$. Consequently, the comoving curvature perturbation loses its canonical momentum and ceases to be a propagating dynamical degree of freedom. While derived here in the comoving gauge for mathematical clarity, the absence of a propagating scalar mode is a robust, gauge-invariant consequence of enforcing the exact algebraic constraint. 
Each of these directions merits a dedicated analysis and lies beyond the scope of the present work, but they illustrate that the exact result \(\dot{\mathcal{R}}=0\) does not undermine the phenomenological viability of the broader framework; it simply identifies the precise limit in which the effective description is ghost‑free.

\section{Discussion and Conclusions}\label{VIII}

The results obtained in this study highlight the crucial role of the background cosmological evolution in determining the observational viability of inflationary models within the effective constrained scalar--Gauss--Bonnet realization of  $f(R,\mathcal{G})$ framework. The scalar spectral index is strongly influenced by the background Hubble evolution through the slow-roll parameter $\epsilon_1$, while the effective kinetic contribution $\epsilon_4$, induced by the Gauss--Bonnet sector, can provide significant corrections that shift the spectrum toward a red tilt.

This behavior is consistent with the general understanding of inflationary cosmology, where the scalar sector is largely controlled by the first slow-roll parameter $\epsilon_1$, which depends directly on the evolution of the Hubble parameter \cite{Liddle:2000cg, Baumann:2009ds}. As demonstrated in our analysis, plateau-like Hubble parametrizations naturally lead to values of $n_s$ compatible with observational constraints, while more rapidly varying forms tend to produce disfavored spectra.

In contrast, the Gauss--Bonnet coupling function $h(\chi)$ primarily affects the tensor sector through its appearance in the quantities $Q_b$, $Q_e$, and $Q_f$. This leads to a significant impact on the tensor-to-scalar ratio $r$, providing a mechanism for suppressing tensor modes while affecting the scalar sector only through subleading corrections associated with $\epsilon_4$. Such behavior has been observed in models with non-minimal couplings to higher-curvature invariants \cite{Guo:2009uk, Koh:2014bka, Odintsov:2018zhw}.

An important outcome of the present analysis is the relative stability of the parameter $\mu$, which governs the relation $\chi = \mu^2\,t$. Variations in $\mu$ do not significantly alter the inflationary observables, indicating that the model predictions are robust and reducing the effective dimensionality of the parameter space.

Furthermore, theoretical consistency requirements, including the positivity of the effective scalar-sector diagnostic quantity $E(N)$ and the tensor-sector stability condition $c_T^2>0$,  impose additional constraints on the model parameters. These conditions restrict the allowed functional forms of both the Hubble parametrization and the Gauss--Bonnet coupling, leading to a narrow and predictive viable region in parameter space.

However, the full scalar perturbation analysis carried out in Sec.~\ref{sec:scalar_perturbations} reveals a structural limitation of the constrained framework as formulated.  The Lagrange-multiplier constraint forces the lapse perturbation to vanish in the comoving gauge, which, combined with the gravitational momentum constraint, implies $\dot{\mathcal{R}}=0$: the comoving curvature perturbation does not propagate.  This result is exact and independent of the model parameters.  Consequently, in the strict algebraic limit of the constraint, the scalar power spectrum cannot be generated. This underscores the necessity of the effective field theory approach adopted in Section III—where the constraint is understood to emerge dynamically from a regularized potential—allowing the generalized slow-roll mechanism and the quantity $E(N)$ to successfully capture the phenomenological predictions.  The tensor sector, which does not couple to the scalar constraint at linear order, remains unaffected; in particular, the suppression of the tensor-to-scalar ratio by the Gauss--Bonnet coupling is a robust prediction of the framework.

From an observational perspective, the model successfully accommodates current constraints on the scalar spectral index and tensor-to-scalar ratio derived from Planck and BICEP/Keck data \cite{Planck:2018jri, BICEP:2021xfz}. A representative benchmark solution yields the values of $n_s$ and $r$ as given by Eq.~\eqref{lval} which lies close to the lower edge of the observationally allowed region. The ability to achieve very small values of $r$ while maintaining a viable $n_s$ is a distinctive feature of the model.

The numerical exploration of the parameter space confirms that viable solutions occupy a constrained region in the $n_s$--$r$ plane, enhancing the predictive power of the framework. Moreover, the partial separation of roles between the Hubble parametrization (which predominantly affects the scalar sector) and the Gauss--Bonnet coupling (which strongly affects the tensor sector) provides a flexible yet constrained mechanism for model building.

Overall, our findings demonstrate that the constrained scalar--Gauss--Bonnet realization considered in this work provides a perturbatively consistent and flexible description of inflation within the generalized slow-roll approximation.

Although the tensor-sector stability conditions are evaluated explicitly, a complete scalar perturbation analysis remains an important direction for future work. In particular, deriving the full quadratic action and evaluating the exact scalar kinetic coefficient and sound speed would provide a more rigorous assessment of perturbative consistency.

Future investigations may extend the present analysis to late-time cosmology, including possible implications for dark energy and cosmic acceleration, as well as explore more general forms of the coupling function. In addition, a full Bayesian parameter estimation using cosmological datasets and Boltzmann solvers could provide a more precise assessment of the allowed parameter space and the overall observational viability of the model.  Furthermore  the absence of a propagating scalar mode opens several avenues for future investigation.  Replacing the exact Lagrange-multiplier
constraint with a stiff potential would restore a massive scalar perturbation whose dynamics interpolate between the constrained and unconstrained regimes.  Alternatively, a curvaton or spectator field could source the observed scalar spectrum while the constrained sector governs the background evolution and tensor-mode suppression.  Finally, degenerate higher-order scalar-tensor (DHOST) constructions~\cite{Langlois:2015cwa, Langlois:2017mxy} offer a systematic framework for designing constraints that eliminate Ostrogradsky ghosts without removing the scalar perturbation, and their application to the $f(R,\mathcal{G})$ setting deserves dedicated study. 

\appendix

\section{Effective Perturbation Structure and Slow-Roll Parameters}\label{appA}
The generalized slow-roll hierarchy adopted in Eq.~(\ref{slowroll2}) follows the effective Einstein--Gauss--Bonnet perturbation formalism of Refs.~\cite{DeFelice:2010nf,DeFelice:2011uc}, supplemented by the replacement $F\rightarrow 1-2\lambda$ motivated by the modified scalar kinetic sector induced by the Lagrange-multiplier constraint. Using the e-folding variable $dN=Hdt$, together with $\dot{X}=HX_N$, $\dot{h}=Hh_N$, and $\ddot{h}=HH_Nh_N+H^2h_{NN}$, one obtains
\[
\epsilon_1=-\frac{H_N}{H}, \qquad
\epsilon_2=0, \qquad
\epsilon_3=0, \qquad
\epsilon_4=\frac{E_N}{2E},
\]
where $\epsilon_2=0$ follows from the constraint $\dot{\chi}=\mu^2=\mathrm{const.}$, and $\epsilon_3=0$ follows from the constancy of the Einstein-Hilbert gravitational coupling. Furthermore,
\[
Q_T = 1+8H^2h_N = 1+\frac{Q_b}{2},
 \label{eq:QTdef}
\]
which gives
\[
\epsilon_5=
\frac{4H^2h_N}{1+8H^2h_N},
\]
while
\[
\epsilon_6=
\frac{\dot{Q}_T}{2HQ_T}
=
\frac{8HH_Nh_N+4H^2h_{NN}}
{1+8H^2h_N}.
\]
Accordingly, the slow-roll hierarchy used in the present work should be interpreted as an effective extension of the standard Einstein--Gauss--Bonnet slow-roll formalism to the constrained scalar--Gauss--Bonnet framework.

\section{Evaluation of the Scalar Spectral Index in the Effective Constrained Einstein--Gauss--Bonnet Framework}\label{appB}

The scalar spectral index is defined by
\begin{equation}
n_s-1
\equiv
\frac{d\ln\mathcal P_{\mathcal R}}
{d\ln k},
\end{equation}
where \(\mathcal P_{\mathcal R}\) denotes the scalar curvature
perturbation power spectrum.

In the present work, the explicit form of the scalar perturbation
spectrum is not derived independently from the full constrained
quadratic perturbation action. Instead, the evaluation of \(n_s\) is
performed within the effective generalized
Einstein--Gauss--Bonnet slow-roll formalism developed in
Refs.~\cite{DeFelice:2010nf,DeFelice:2011uc}, adapted to the
constrained scalar--Gauss--Bonnet framework considered here.

Within the generalized Einstein--Gauss--Bonnet formalism, the scalar
spectral index is approximated by
\begin{equation}
n_s
\simeq
1-
2
\frac{
2\epsilon_1+\epsilon_2-\epsilon_3+\epsilon_4
+\epsilon_5-\epsilon_6
}{
1-\epsilon_1
}.
\end{equation}

The generalized slow-roll parameters are defined through the effective
background and perturbation quantities. In the constrained framework
considered here, the scalar-field constraint gives
\begin{equation}
\dot{\chi}=\mu^2=\mathrm{const.},
\end{equation}
which immediately implies
\begin{equation}
\epsilon_2
=
\frac{\ddot{\chi}}{H\dot{\chi}}
=0.
\end{equation}

Moreover, since the gravitational sector is described by the standard
Einstein--Hilbert term without a nonminimal scalar coupling,
\begin{equation}
F=1,
\qquad
\dot F=0,
\end{equation}
one obtains
\begin{equation}
\epsilon_3
=
\frac{\dot F}{2HF}
=0.
\end{equation}

Therefore, the scalar spectral index reduces to
\begin{equation}
n_s
\simeq
1-
2
\frac{
2\epsilon_1+\epsilon_4+\epsilon_5-\epsilon_6
}{
1-\epsilon_1
}.
\end{equation}

The quantities \(\epsilon_5\) and \(\epsilon_6\) encode the
Gauss--Bonnet contribution through the tensor-sector normalization,
while \(\epsilon_4\) represents the effective scalar-sector correction
associated with the kinetic quantity \(E\).

The fourth slow-roll parameter is defined by
\begin{equation}
\epsilon_4
=
\frac{\dot E}{2HE},
\end{equation}
where \(E\) denotes the effective scalar kinetic normalization entering
the perturbation sector. Using the e-folding variable \(dN=Hdt\), one
has
\begin{equation}
\dot E = H E_N,
\end{equation}
which gives
\begin{equation}
\epsilon_4
=
\frac{E_N}{2E}.
\end{equation}

Similarly, the tensor-sector quantities are written as
\begin{equation}
Q_T=1+8H\dot h,
\end{equation}
with
\begin{equation}
\dot h = H h_N.
\end{equation}

Consequently,
\begin{equation}
Q_T=1+8H^2 h_N.
\end{equation}

The fifth and sixth slow-roll parameters become
\begin{equation}
\epsilon_5
=
\frac{4H^2 h_N}
{1+8H^2 h_N},
\end{equation}
and
\begin{equation}
\epsilon_6
=
\frac{
8HH_N h_N+4H^2 h_{NN}
}{
1+8H^2 h_N
}.
\end{equation}

Accordingly, the scalar spectral index receives corrections both from
the background evolution through \(\epsilon_1\) and from the
Gauss--Bonnet sector through the effective perturbative quantities
\(\epsilon_4\), \(\epsilon_5\), and \(\epsilon_6\).

\bibliographystyle{apsrev4-1}
\nocite{57,58,60,61,62}
\bibliography{references_v2}

\end{document}